\documentclass[AMA, LATO1COL]{WileyNJD-v2}

\usepackage{multirow}
\newcolumntype{L}[1]{>{\raggedright\let\newline\\\arraybackslash\hspace{0pt}}m{#1}}
\newcolumntype{C}[1]{>{\centering\let\newline\\\arraybackslash\hspace{0pt}}m{#1}}
\newcolumntype{R}[1]{>{\raggedleft\let\newline\\\arraybackslash\hspace{0pt}}m{#1}}

\articletype{Article Type}%

\received{}
\revised{}
\accepted{}

\begin{document}

\title{Misaligned Values in Software Engineering Organizations}

\author[1]{Per Lenberg*}
\author[1]{Robert Feldt}
\author[2]{Lars G{\"o}ran Wallgren Tengberg}

\address[1]{\orgdiv{Department of Computer Science and Engineering}, \orgname{Chalmers University}, \orgaddress{\city{Gothenburg}, \country{Sweden}}}
\address[2]{\orgdiv{Department of Psychology}, \orgname{Gothenburg University}, \orgaddress{\city{Gothenburg}, \country{Sweden}}}

\corres{*Corresponding author. \email{perle@chalmers.se}}
\abstract[Summary]{
The values of software organizations are crucial for achieving high performance; in particular, agile development approaches emphasize their importance. Researchers have thus far often assumed that a specific set of values, compatible with the development methodologies, must be adopted homogeneously throughout the company. It is not clear, however, to what extent such assumptions are accurate.


Preliminary findings have highlighted the misalignment of values between groups as a source of problems when engineers discuss their challenges. Therefore, in this study, we examine how discrepancies in values between groups affect software companies' performance.

To meet our objectives, we chose a mixed method research design. First, we collected qualitative data by interviewing fourteen (\textit{N} = 14) employees working in four different organizations and processed it using thematic analysis. We then surveyed seven organizations (\textit{N} = 184). Our analysis indicated that value misalignment between groups is related to organizational performance. The aligned companies were more effective, more satisfied, had higher trust, and fewer conflicts.

Our efforts provide encouraging findings in a critical software engineering research area. They can help to explain why some companies are more efficient than others and, thus, point the way to interventions to address organizational challenges.
}


\maketitle
\section{Introduction}
\label{sec1}
The introduction of agile approaches marks a paradigm shift for the software industry; they have affected development methods by advocating for adaptive planning, evolutionary development, early delivery, continual improvement, and responsiveness to change~\cite{abrahamsson2017agile}. They have also influenced how software companies organize their work by shifting the focus from the individual developer, instead highlighting the significance of teams, collaboration, and communication~\cite{highsmith2002agile,cockburn2006agile, lenberg2015behavioral}. As a direct result, in modern software organizations, the team has replaced the individual as the most critical entity.

In addition to having significant effects on how software companies organize and what methods they use, agile approaches have also shaped the companies' cultures and organizational value structures. The agile manifesto~\cite{beck2001manifesto}, which is the founding document for agile approaches, emphasize the significance of organizational values. Therefore, over the past 15 years, researchers have explored the intricate link between agile methods and organizational values or culture~\footnote{Culture relates strongly to values. It is, however, a broader more all-inclusive term that covers more aspects of organizational life (see Section ~\ref{lbl_organizational_culture})}. These studies have often relied on an assumption of compatibility or fit~\cite{iivari2011relationship}. Using the cultural dimensions identified by Hofstede (clan, democratic, hierarchical, and disciplined)~\cite{hofstede2001organizations}, Siakas and Siakas~\cite{siakas2007agile} identified the democratic culture dimension as the most suitable for an agile approach. Strode et al.~\cite{strode2009impact} investigated the relationship between the competing values framework (CVF)~\cite{denison1991organizational} and the agile XP method. Data extracted from nine projects indicated most consistently significant associations between XP and the collaborate dimension of CVF.

Even if studies have repeatedly recognized that successful agile adoption entails reforms to the existing value foundation~\cite{chow2008survey, chandra2010identifying, tolfo2011agile, hamid2015factors, dikert2016challenges}, such conclusions are not uncontested. For example, Robinson and Sharp~\cite{robinson2005organisational} argue that, due to their innate flexibility, agile approaches can thrive in a variety of cultural settings, while Siakas and Siakas~\cite{siakas2007agile} suggest that organizational agility should be regarded as a culture of its own. We note, however, that outside of the framing of agile transitions, software engineering studies exploring organizational values have been scarce.

Given that social science research has repeatedly recognized the critical role that values play in various facets of organizational life~\cite{belias2014organizational, schneider2013organizational, hartnell2011organizational, leidner2006review}, we believe that broader and more extensive insights on organizational values are likely to be beneficial for software engineering. A social science strategy that is related to organizational values and used to describe and explain organizational behavior is alignment theory~\cite{quiros2009organizational}. It draws on the assumption that in order to achieve effectiveness, the organizational entities must be directed and structured so that they are suited to each other~\cite{nadler1988strategic}. Research has, for example, shown that values alignment fosters collaboration and could be a proactive approach to conflict management~\cite{lynn2007literature}.

One of our previous studies also demonstrated the potential usefulness of value alignment in the software engineering context~\cite{lenberg2018used}. The findings indicated that discrepancies in shared values between organizational groups adversely affected performance. In the present deductive study, we aimed to test these initial findings further and thereby to delineate and add support to a between-group value misalignment theory. Our primary objective was, therefore, to \emph{examine how discrepancies in values between organizational groups affect software companies' performance}. To the best of our knowledge, the effects of between-group value misalignment have not previously been explored within the software engineering context.

In addition to the relatively narrow primary objective, we also aimed to extend the knowledge of organizational values more broadly using an exploratory research approach. Accordingly, our secondary objective was to \emph{gain general insights into organizational values and how they affect behaviors and performance in software companies}.

\begin{figure}
\centering
\includegraphics[width=0.90\textwidth]{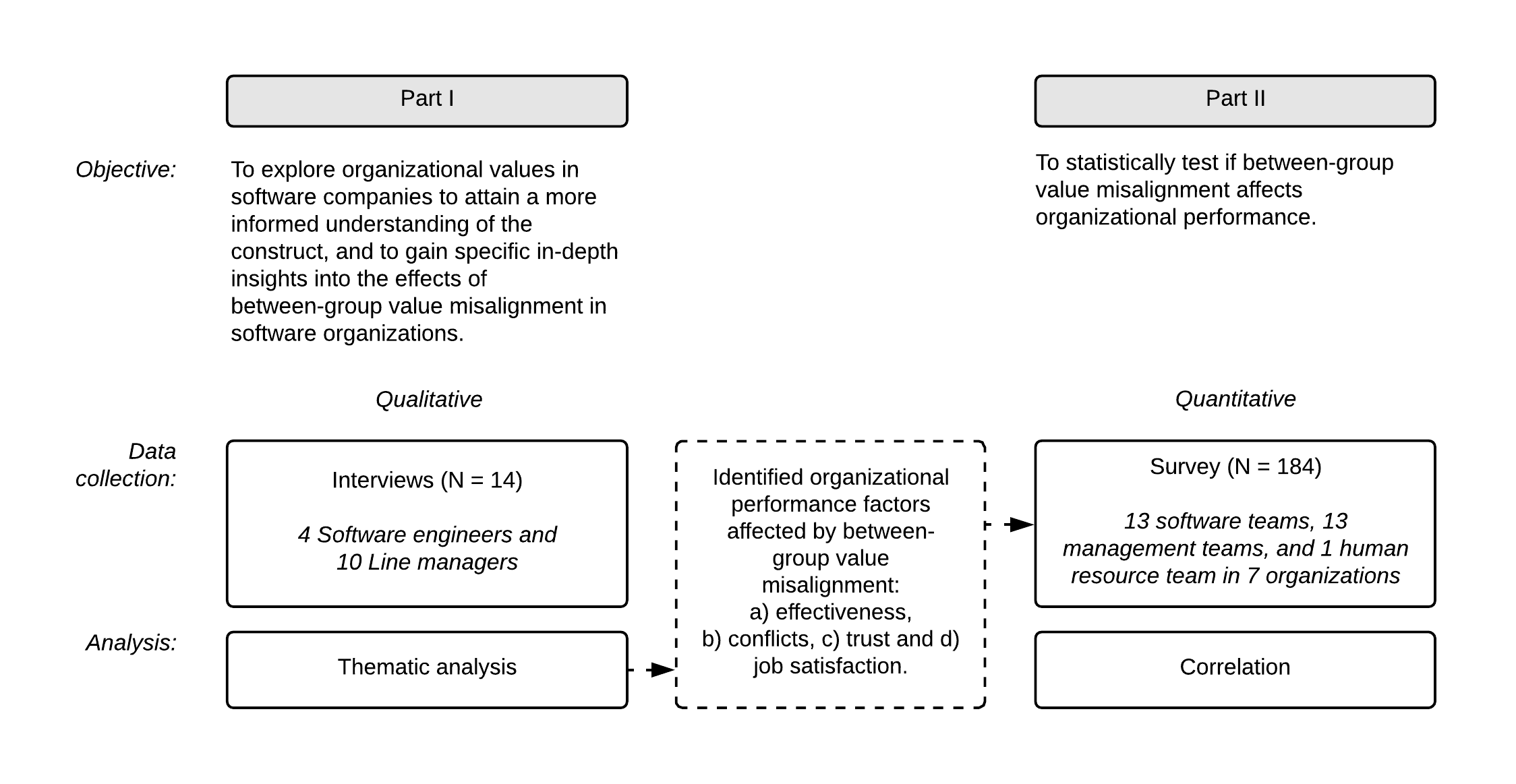}
\caption{An overview of the study.}
\label{fig_method_overview}
\end{figure}

To meet these objectives, we selected a mixed method research design~\cite{creswell2013research} (see Figure~\ref{fig_method_overview}). First, we collected qualitative data by interviewing 14 (\textit{N} = 14) employees working in four different software engineering organizations. We strove to gain insights as to the effects of between-group value misalignment and what organizational performance factors that were affected. The data were processed using thematic analysis~\cite{braun2006using}. Then, to statistically test if value misalignment related to the performance factors that we had identified in the qualitative analysis, we surveyed seven organizations (\textit{N} = 184). In the questionnaire, we utilized the CVF~\cite{denison1991organizational} to estimate organizational values and as the basis for calculating the between-group value misalignment. 

\begin{figure}
\centering
\includegraphics[width=0.65\textwidth]{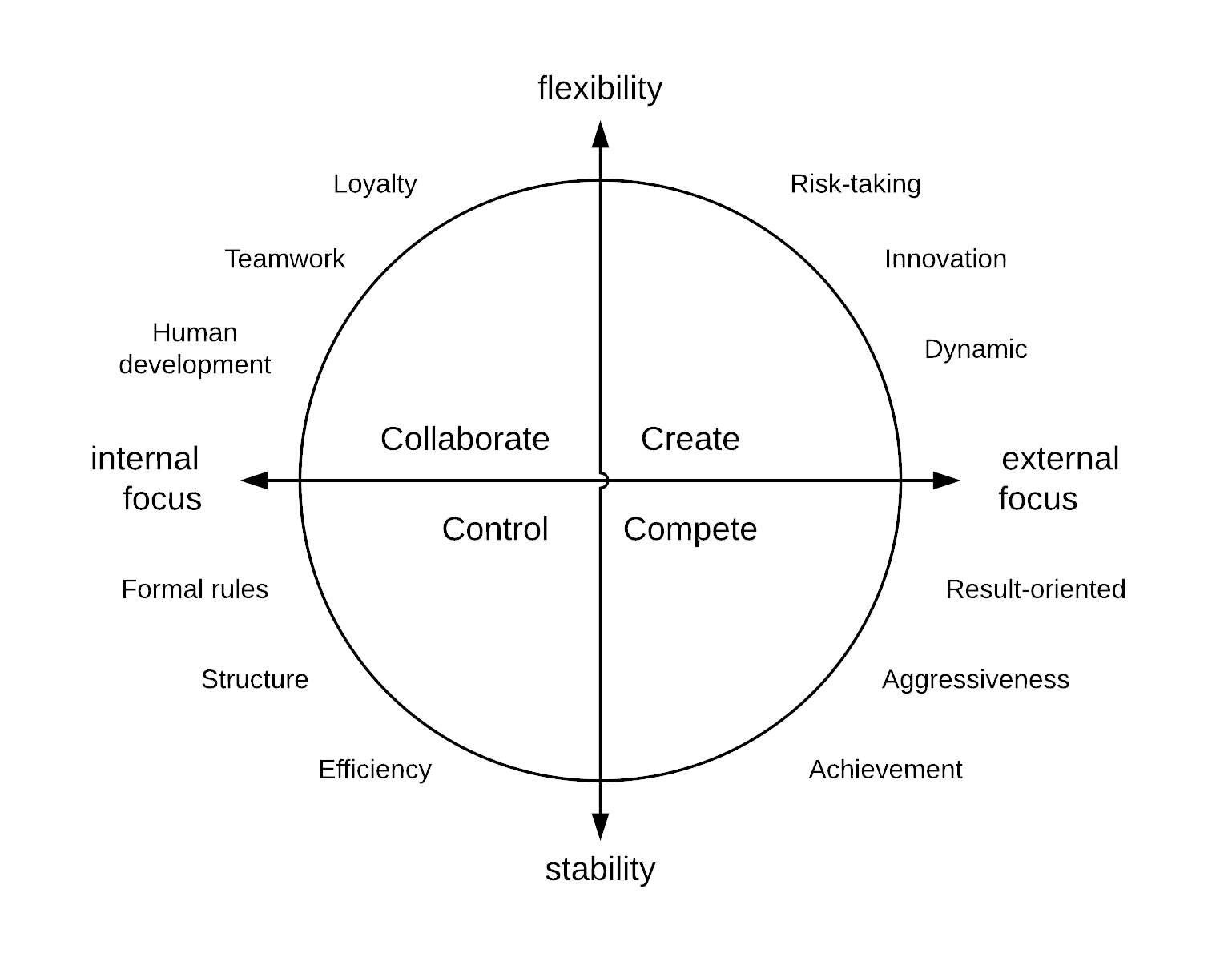}
\caption{An overview of the competing values framework~\cite{denison1991organizational}.}
\label{fig_cvf_figure}
\end{figure}

Taken together, we argue that our research contributes by demonstrating the effects of value misalignment between groups in software development. That is important for software engineering organizations since they, almost exclusively, organize their work in groups or teams. Our efforts also contribute by providing more general insights on values in software engineering organizations. This is relevant considering that such research has had a limited focus, primarily exploring the fit between specific values and agile approaches~\cite{iivari2011relationship}.

In the next section (Section~\ref{sec_back}), we give further background information on organizational value and other similar concepts and provide an overview of organizational value research in software engineering. We then describe the methods, analysis, and results of the two parts of our mixed research design study (see Section \ref{section_study_a} and \ref{section_study_b}). Finally, the aggregated findings are discussed (\ref{section_discussion}) and concluded (\ref{section_conclusion}).

\section{Background}
\label{sec_back}
In the following sections, we provide information that we deemed relevant to the study. First, we define the construct organizational values and frame our research in a contextual and historical setting by providing an overview of the research on organization values in software engineering. We also briefly describe other similar constructs and, finally, summarize the literature on organizational alignment.

\subsection{Organizational values}
The initial research on values focused on the personal values of individuals~\cite{malbasic2015balance}. Personal values define preferences and thus reflect what is essential to individuals~\cite{lynn2007literature, hultman2002balancing}. In an organization, individual values guide the employees' private decisions and actions, while \emph{organizational values} provide norms that specify how they should behave and how organizational resources should be allocated~\cite{edwards2009value}. According to a review of \emph{organizational values} by Stavru~\cite{stavru2012organizational}, the most studied constructs concerning \emph{organizational values} are organizational commitment, employee retention, and well-being.

The literature shows multiple definitions of the \emph{organizational value} construct. Virtually all of them, however, acknowledge that the construct operates as a guide to the decision-making process and that it is used to evaluate individual and organizational actions and states~\cite{stavru2013we}. In this study, we use a definition adopted from Enz~\cite{enz1988role} that describes \emph{organizational values} as \emph{beliefs, a group of persons, express by preference in the context of identifying desirable courses of action and goals}.

The research into \emph{organizational values} in software engineering has mostly been conducted within the framing of the \emph{organizational culture} construct. For the past 15 years, the focus of the research has primarily been to explore the relationship between agile approaches and culture, where the studies have often relied on an assumption of compatibility or fit~\cite{iivari2011relationship}. For example, using the cultural dimensions identified by Hofstede (clan, democratic, hierarchical and disciplined)~\cite{hofstede2001organizations}, Siakas and Siakas~\cite{siakas2007agile} identify the democratic culture type as the most suitable for an agile approach. Strode et al.~\cite{strode2009impact} investigated the relationship between CVF~\cite{denison1991organizational} and the agile XP method. Data extracted from nine projects indicated most consistently significant associations between XP and the collaborate culture. The CVF was also used as a basis in work by Iivari and Iivari~\cite{iivari2011relationship}. The authors, somewhat in conflict with the study by Strode, suggest that all culture CVF orientations (except control) favor agile methods.

Moreover, Tolfo et al.~\cite{tolfo2011agile} explored a view of organizational culture in three levels as a theoretical framework to allow early detection of problems. The authors note that many facilitators of or obstacles to the adoption of an agile approach can be hidden in the lower, latent, levels of the culture. Although these studies emphasize the importance of recognizing the complicated interplay between agile methods and organizational culture, they do not provide hands-on guidance as to how to introduce and adopt agile methods into the organizational culture.

The majority of the studies have recognized that successful agile adoption entails reforms to the existing value foundation~\cite{chow2008survey, chandra2010identifying, tolfo2011agile, hamid2015factors, dikert2016challenges}. Still, the research is not unanimous. Robinson and Sharp~\cite{robinson2005organisational} suggest that due to their innate flexibility, agile approaches can thrive in a variety of cultural settings, while Siakas and Siakas~\cite{siakas2007agile} propose that organizational agility should be regarded as a culture of its own. 

Even if agile approaches have framed the primary focus of organizational values research, there are exceptions. A study by Shih and Huang~\cite{shih2010exploring} explored the relationship between software process improvement (SPI) and organizational culture. Using the CVF, they conclude that an SPI deployment was made possible primarily by a control culture that emphasizes procedure, order, and stability. Furthermore, two studies by Mathew~\cite{mathew2007relationship} and Lavallee and Robillard~\cite{lavallee2015good} using qualitative research approaches provide fascinating, yet initial, results that strengthen the link between organizational culture and software quality. A literature review by Purna~\cite{purna2011soft} also adds support to this relationship.

\subsubsection{Related constructs}
According to Schneider et al.~\cite{schneider2013organizational}, the complexity of organizational behavior calls for various constructs used for description and analysis of organizational life. In the literature, the following three constructs relate to and partially overlap with \emph{organizational values}: \emph{organizational culture}, \emph{organizational climate}, and \emph{organizational identity}. These constructs are detailed in the following sections.

\paragraph{Organizational culture}
\label{lbl_organizational_culture}
The \emph{organizational culture} construct has its conceptual and methodological basis in anthropology~\cite{schneider2013organizational}. Its natural unit of analysis is thus the collective, whereas differences among individual employees tend to be of less interest. Compared to \emph{organizational values}, it is a broader, more all-inclusive, term that may cover almost every aspect of organizational life, for example, basic assumptions and beliefs, values, models of behavior, rituals, practices, symbols, heroes, artifacts, and technology~\cite{iivari2011relationship, hartnell2011organizational, schneider2013organizational}.

\begin{figure}
\centering
\includegraphics[width=0.55\textwidth]{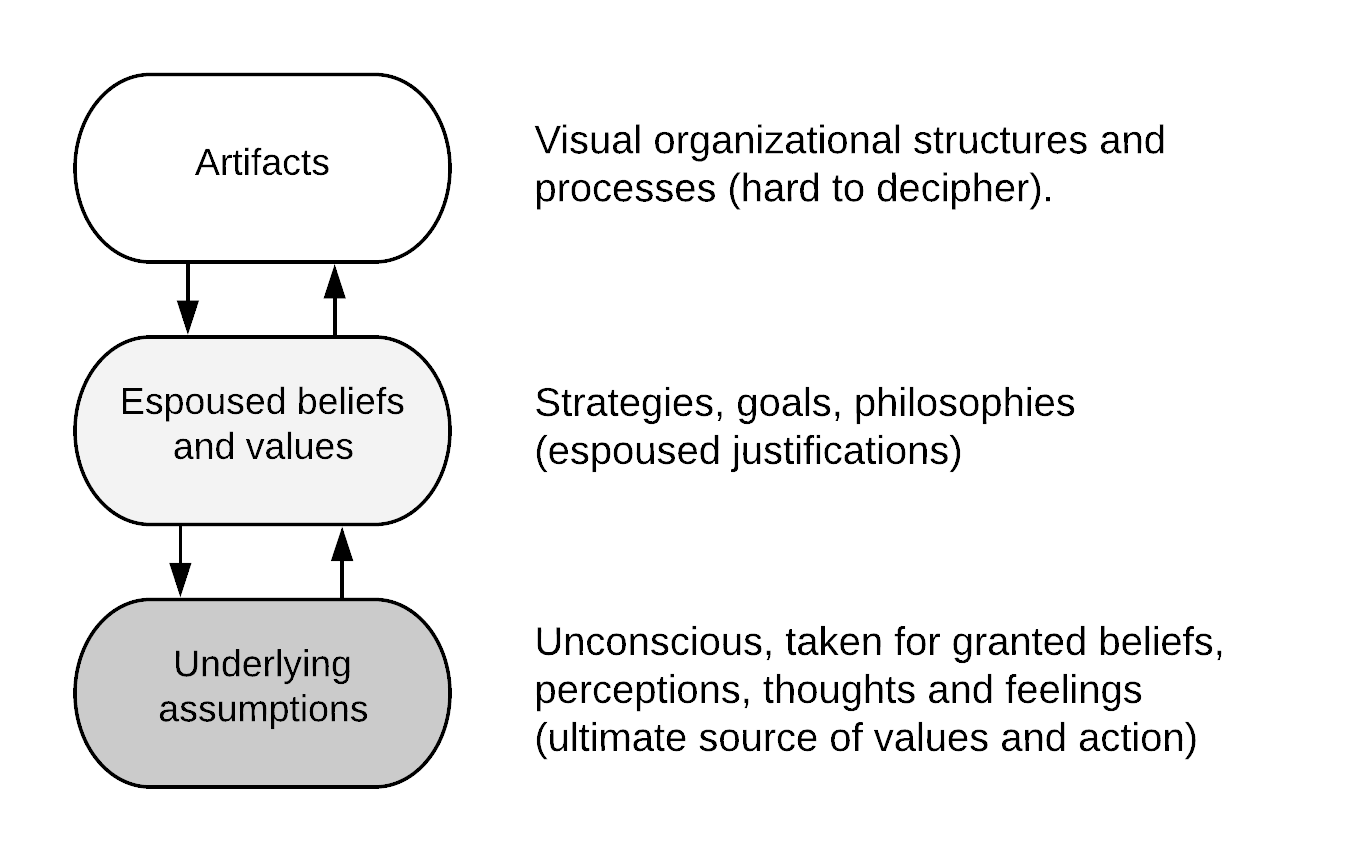}
\caption{An overview, adopted from Schein~\cite{schein2010organizational}, of the three layers in his model.}
\label{fig_schein}
\end{figure}

There seems to be agreement among researchers that the \emph{organizational culture} construct has several levels~\cite{hofstede1990measuring, schein2010organizational, schneider2013organizational}. One of the most prominent and well-known models used to capture these levels is the culture framework defined by Schein~\cite{schein2010organizational}. His model consists of three levels: artifacts, espoused beliefs and values, and underlying assumptions (see Figure~\ref{fig_schein}).

Artifacts represent phenomena that are visible, including, for example, rituals, working routines, language, myths, ways of dressing, and the organization of workspace. These are the most readily accessible to outsiders, but also the most ambiguous concerning the underlying meaning they may represent. Thus, although many artifacts may look the same across organizations, the meaning(s) ascribed to them may be entirely different~\cite{schneider2013organizational}.

The intermediate level consists of espoused values or the values that are reported by management as being core to the organization. These may or may not be reflected in the employee's actual organizational behavior. According to Schein, the leadership should have strong influential skills in order to make such values acceptable by employees. The values allow organizational members to interpret signals, events, and issues that guide behavior.

The final level (underlying assumptions) indicates why employees go about their day-to-day work lives as they do, and are frequently so ingrained that they cannot necessarily be easily articulated, requiring in-depth interviewing to illuminate them. Underlying assumptions develop over time when members of a group create strategies to face problems.

According to Schneider et al.~\cite{schneider2013organizational}, researchers have conceptualized the culture construct in two way: by focusing on culture either as something an organization \emph{has} or something an organization \emph{is}. The research done from the former perspective is usually comparative and uses quantitative surveys to uncover the attributes that differentiate more efficient organizations from less efficient ones. As for the latter approach, the researchers' goal is more exploratory, aiming at uncovering fundamental assumptions and root metaphors. Such research approaches tend to be qualitative and inductive in order to report how insiders experience their organizations.

\paragraph{Organizational climate}
A closely related construct to \emph{organizational culture} is \emph{organizational climate}. The two constructs have been used interchangeably in research studies, and there is no definite distinction between the two~\cite{mclean2005organizational, zohar2012organizational}. While \emph{organizational culture} is, as stated above, the shared beliefs and assumptions about the organization's expectations and values, \emph{organizational climate} is the shared perceptions and attitudes about the organization. It may be defined as the meaning attached to the policies, practices, and procedures employees experience and the behaviors they observe being rewarded~\cite{schneider2013organizational, ostroff2003organizational}.

Schein~\cite{schein2010organizational} suggests that climate provides behavioral evidence for the culture, such that those behaviors form the bases for employees' conclusions about the values and beliefs that constitute the organizational culture.

It is also worth noting that while \emph{organizational culture} has naturally been considered a collective construct, \emph{organizational climate} research has struggled with a unit of analysis issue (i.e. whether climate is an individual experience construct or an organizational attribute). Today, the vast majority of the climate research conducted is collective, but there are still a few exceptions.

\paragraph{Organizational identity}
The construct of \emph{organizational identify} was defined by Albert and Whetten in 1985 and later clarified in 2006~\cite{albert1985organizational, whetten2006albert}. They define \emph{organizational identity} as a set of statements that employees believe are central, distinctive, and enduring to their organization. Central means that the statements should include features that are critical to the organization, while distinctive emphasizes that the identity statement should be able to distinguish the organization from others. Finally, `enduring' indicates that the identity statements are stable in the organization over time. 
According to this definition, an identity statement is collectively and cognitively held by organization members to answer questions such as `Who are we?' or `What do we want to be?'. \emph{Organizational identity} often attempts to apply sociological and psychological concepts and theories about identity to organizations.

The relationship between identity and culture has long been debated among academics, and the two concepts clearly overlap. Drawing upon Mead's framework~\cite{mead1934mind}, Hatch and Schultz~\cite{hatch2002dynamics} propose a dynamic model to illustrate the relationship between organizational identity, culture, and image. According to their model, employees express their understanding of their organizational culture through identity, which affects the perception of others outside the organization about the organization. The outsiders' perception, or organizational image, in turn, affects the organizational identity, which again is reflected in the central elements of the organizational culture~\cite{mujib2017organizational}. According to this model, \emph{organizational identity} has (in contrast to \emph{organizational values}, \emph{organizational culture}, and \emph{organizational climate}) a component related to others' perceptions of an organization.

\subsection{Organizational alignment}
Alignment theory - a relatively recent approach used to explain organizational life - aims at the need for alignment among the cultural, structural, and strategic components of an organization~\cite{quiros2009organizational}. The approach is based on the assumption that in order to achieve effectiveness, all of the organizational entities must be directed and structured so that they are suited to each other. A seminal framework designed for studying alignment was compiled by two pioneers of the field, Nadler and Tushman~\cite{nadler1988strategic}.

The literature distinguishes between two types of organizational alignment: vertical and horizontal. Vertical alignment refers to the congruence of strategies, goals, and objectives between various hierarchical levels. Horizontal (also known as lateral) alignment refers to the coordination of efforts across an organization. It is related to the consistency of decisions across entities, so that activities across, for example, marketing, operations, HR, and other functions support one another~\cite{kathuria2007organizational}.

The fact that effective companies tend to have shared and aligned values has sizable empirical support~\cite{detert2000framework,denison1991organizational, burnes2011success, stavru2013we}. Research has shown that values alignment fosters collaboration and could be a proactive approach to conflict management~\cite{lynn2007literature}. However, a majority of the studies on alignment have been related to the alignment (or congruence) of values between the organization and individual employees, also referred to as person-to-organization fit~\cite{hultman2002balancing, burnes2011success, edwards2009value, stavru2013we}. Clear~\cite{clear2010exploring} has also conducted initial research into cultural fit and virtual teams. In the paper, the author outlined a model emphasizing that the cultural construct is complex and includes several layers that organizations need to consider.

To the best of our knowledge, only a few studies exist that have explored the alignment of values between groups in a software engineering context. Huang et al.\cite{huang2003dangerous} investigated the link between inconsistencies in organizational subcultures and the introduction of component-based software development methods. They stressed that misalignment of values among subcultures hindered the information sharing and collaboration needed to integrate a method effectively. Furthermore, in 2012, Stavru outlined a study designed to explore the relationship between organizational values and the deployment of agile methods~\cite{stavru2012organizational}. However, the results of that study (if it was conducted) are yet to be published.

\section{Part I - Exploring the effects of value misalignment}
The purpose of this study was dual. First, we aimed to broadly explore organizational values in software companies to attain a more informed understanding of the construct. Second, we strove to gain specific in-depth insights into between-group value misalignment in software organizations and how it affect behaviors and performance.

\label{section_study_a}
\subsection{Method}
Since we sought to explore organizational values and gain in-depth insights into value misalignment, we undertook a qualitative study in which we collected data using interviews. The following sections present an overview of the participating companies, our sample, the data collection process, and our analysis method.

\subsubsection{Company overview}
\label{company_overview}
We collected industry data from seven departments in six companies. For simplicity reasons, we will, in this study, refer to the departments as organizations. The companies were international (employees in multiple countries) and large (>3000 employees). A brief description of the organizations is found in Table~\ref{table:companies}. It presents the number of employees, number of interviewees in part I of this study, number of respondents and teams in part II (survey), a description of the department, and, finally, what agile approach they were using.

All of the participating organizations claimed to use an agile development approach. Drawing on the guidelines from Gren, Torkar, and Feldt~\cite{gren2015prospects} and using the knowledge we gained through the interviews (see Section~\ref{section_study_a}), we deemed that the organizations were all on level two or three of the Agile Adoption Framework developed by Sidky~\cite{sidky2007disciplined}.

We selected the organizations using convenience sampling by contacting 12 organizations that we had contacted previously. Of these, four agreed to participate in the interviews and the surveys, three were only willing to conduct the surveys, three organizations did not respond, and two declined due to high workload. The seven participating organizations were all located in Sweden.

\begin{table}
\footnotesize
\setlength\extrarowheight{2pt}
\centering
\begin{tabular}{ C{.03\textwidth}  C{.07\textwidth}  C{.07\textwidth} C{.07\textwidth} L{.5\textwidth} C{.07\textwidth}  } \hline
Org. Id. & No. of employee & No. of interviewees & No. of respondents & Description & Agile framework \\ \hline
A & 800 & 4 & 32 (5) & A department within a large engineering company (the same company as department C). The department had employees in Sweden and in the United States. Among the 800 employees, roughly 120 were software engineers developing both low-level components and high-level applications used in the department's products. & Scrum \\ 
B & 1000 & 0 & 35 (4) & An engineering company that had existed for more than 50 years. It manufactured complete products, not only software. The managers were thus not only responsible for software engineers. They were also responsible for units that designed, produced, and developed mechanical constructions. The company had roughly 80 software engineers organized in 12 teams, developing low-level software components. & Scrum \\
C & 120 & 3 & 22 (4) & A department within the same company as A. However, the two departments were separated both geographically and hierarchically. The common denominator was the CEO of the company. The department, which had almost one 100 software engineers, developed large customized business-to-business systems that included both hardware and software. & SAFe \\
D & 160 & 3 & 31 (4) & An in-house software consultant company. The department that participated in our study had approximately 160 software engineers working in roughly 40 teams. As a consultant company, they developed both low-level components and high-level applications. & Scaled scrum \\
E & 150 & 4 & 23 (4) & A department with 50 software engineers within a large system development company that developed low-level software components used by the other departments in the company. & SAFe \\
F & 150 & 0 & 19 (3) & A department that specialized in developing advanced software solutions used by a large system development company. All employees were software engineers. & SAFe \\
G & 100 & 0 & 22 (3) & A small department with roughly 1000 software engineers within the large IT consultant company that had both in-house software teams and regular consultants. The participating teams were in-house consultants. & Scrum \\ \hline
\end{tabular}
\caption{An overview of the seven departments that participated in the study.}
\label{table:companies}
\end{table}

\subsubsection{Sample}

The sample population of this sub-study was employees in the four organizations (i.e. A, C, D, and E) described in Table~\ref{table:companies}. In total, 14 employees consisting of four women and ten men aged from 29 to 60 years with an average of 43 years participated. All participants had worked in their respective organizations for more than five years and had held their current position for at least two years.

\begin{figure}
\centering
\includegraphics[width=0.85\textwidth]{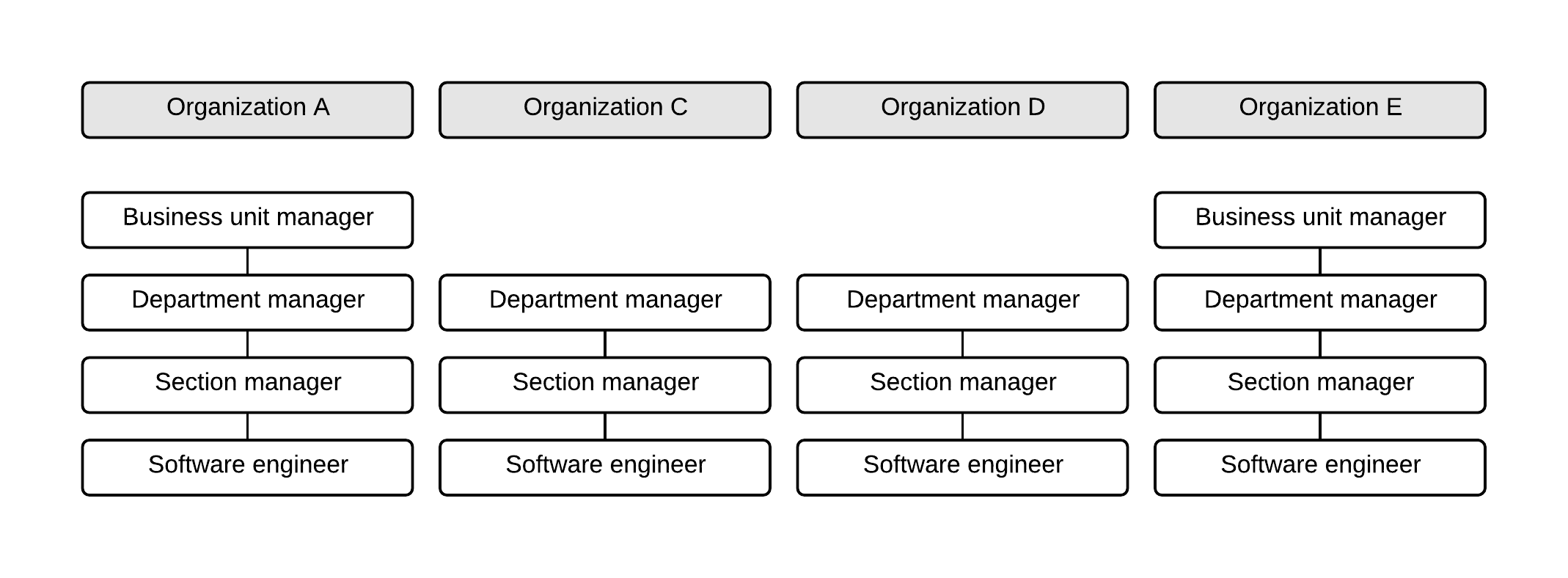}
\caption{An overview of the interviewees.}
\label{fig:stda_interviewees}
\end{figure}

As can be seen in Figure~\ref{fig:stda_interviewees}, in two of the organizations, four employees participated, and in the other two organizations, three employees participated. In each organization, the participating employees had different hierarchical positions: one software engineer (member of an agile team), one section manager, one department manager, and, for two of the included organizations, one business unit manager.

The participants from respective organizations were hierarchically in line, which constrained the selection process. The upper management provided a list of managers that reported directly to them. From that list, we randomly selected three managers that we contacted by email. We chose to include the first manager that responded. The selection of the other participants followed the same process. Roughly 50\% of the participants we contacted by email replied. Of these, all were willing to participate.

\subsubsection{Materials}
\label{lbl_materials}
We used semi-structured interviews since this format allows the interviewer to ask follow-up questions to gain in-depth knowledge and explore concepts ad hoc through dialog with the interviewee~\cite{smith2007qualitative, de2001research}. We used the following primary questions as a basis for the interviews:

\begin{enumerate}
\item[i] Could you please describe your job at [company]?
\item[ii] Could you describe a satisfactory working day for you?
\item[iii] What constitutes prosperous behavior in your organization?
\item[iv] What behavior harms your organization?
\end{enumerate}

The primary questions were designed to expose the organizational behaviors that the interviewees and their respective organizations found acceptable and prosperous and vice versa. However, these questions were not directly related to the purpose of the interviews (i.e. to gain insights into organizational values and value misalignment). The researcher instead used follow-up questions to gravitate the conversations toward the topics of primary interest to this study. For example, if the interviewee mentioned that there were trust issues between organizational units, the researcher asked questions to uncover the underlying causes (e.g. `Why don't you trust the members in Unit X?' and `How does the lack of trust affect your work?'). We chose this method to try to ensure that the dialogue was grounded in real experiences and actual opinions. The first two questions related to the behavior of the participants, while the last two questions focused on the participants' experiences of behavior in the organization as a whole.


\subsubsection{Procedure}
The interviews were conducted on the respective organizations' premises during a period of six weeks from October to December of 2017. The sessions were performed face-to-face by a single researcher in a separate conference room without the risk of disturbance by others. The sessions were held in Swedish and lasted, on average, 45 minutes. All sessions were audio recorded.

The interviews began with the interviewer informing the interviewee that participation was voluntary, and that he or she could end the interview at any time. The interviewee was told that she or he did not represent any team, group, unit, or organization and that instead the interviewer's primary interest was the interviewee's true opinions and experiences. Moreover, the interviewee was informed that his or her interview would be confidential and that the raw and analyzed data would not be distributed to anyone. The interviewer also presented the objectives of the study and provided some background information to the research area and asked the interviewee if he or she had any questions regarding the interview procedure or the background materials.

The interviewer then began the formal interview based on the questions outlined in Section~\ref{lbl_materials}. At the end of the interview, when the interviewer had no more follow-up questions, the interviewees were asked if they had any finally comments or reflections and if they felt that their true opinions had surfaced.
To verify the procedure and the interview questions, we conducted one pilot interview. Based on its results, we saw no reason to alter the procedure or questions. The pilot interview was, however, not included in the analysis.

\subsubsection{Analysis}
To process the data, we chose to use thematic analysis based on the guidelines outlined by Braun and Clarke~\cite{braun2006using}. Thematic analysis is considered a flexible method that can be used for reporting experiences and is a method that we, the authors, have previously used~\cite{lenberg2015human}. A single author (i.e. the first author) conducted the analysis. The result of each phase (described below) was, however, discussed with the co-authors.

In the analysis, which was based on realism underpinnings~\cite{scotland2012exploring}, we did not aim to provide a rich thematic description of the entire data. Our ambition was, instead, to provide a detailed and nuanced account for a group of themes related to the dual aim of this sub-study. We thus limited our themes to relate to organizational values in software companies and, more specifically, to the causes and consequences of value misalignment.

Our ambition was to create insights that go further than the participants own understanding. To accomplish that, we chose a latent analysis depth, meaning that we tried to go beyond the semantic content of the data, and examine the underlying ideas and assumptions that are shaping the semantic content~\cite{braun2006using}. Also, we used an inductive, data-driven, approach (i.e. we drew theoretical and general conclusions from the data).

The analysis included the following phases:
\subparagraph{Data familiarization} We transcribed the interviews, after which we read and reread them to obtain a general understanding and idea of the content. This process allows any new information obtained to affect the interpretation of the general content~\cite{kvale2014kvalitativa}. We analyzed the data organization by organization, starting with the top manager. We hoped that this strategy would facilitate the comparison of individual experiences and opinions as well as the comparison between organizations. 
\subparagraph{Coding} We then identified initial codes from the transcripts. A code identifies a feature of the data that appears of interest to the aim of the study (i.e. it related to organizational values or value misalignment). The initial codes were then grouped in a thematic map~\cite{cruzes2011recommended}.
\subparagraph{Comparison sorting} As a pre-step to the theme identification phase, we roughly sorted the codes based on differences of opinions. This step created pairs of code clusters, where the codes of one of the clusters supported one opinion while the other cluster included codes supporting the other viewpoint.
\subparagraph{Theme identification} Next, we categorized the codes into sub-themes and themes, and analyzed the relationships between them~\cite{smith2015qualitative} using the thematic map as well as surrounding data. This step is a reductive abstracting process, aimed at finding an integrated, meaningful conceptual pattern overlaying the data~\cite{hellstrom2001affecting}. 
\subparagraph{Theme defining and naming} In the final step, we reread the data and assigned citations that illustrated the final themes. The citations were primarily selected based on the relevance of the theme; however, we also wished for citations to cover a significant part of the collected data.

\subsection{Results}
Our analysis resulted in three themes. These are, together with their respective sub-themes, presented in Table~\ref{table_themes}. The first theme, labeled \emph{the dark side of agile}, relates to the adverse effects caused by the value structure that, according to the interviewees, tends to follow the introduction of the agile approaches. The second theme, labeled \emph{triggers and winds of change}, summarizes the main factors that will form the value structure in future software organizations. The third and final theme, labeled \emph{value misalignment}, relates to the causes and effects of alignment, and misalignment, between groups in software engineering organizations.

\begin{table}
\footnotesize
\setlength\extrarowheight{2pt}
\centering
\begin{tabular}{ L{.2\textwidth} L{.3\textwidth} L{.4\textwidth}} \hline
Theme & Sub-theme & Summary \\
\hline
\multirow{3}{*}{1. The dark side of agile} & 1.1 Political correctness & 
\multirow{3}{.4\textwidth}{The theme relates to the potential adverse effects caused by the change in value structure that follow the introduction of the agile approaches.} \\
& 1.2 One ambiguous method to rule them all & \\
& 1.3 Agile as trademark & \\
\multirow{3}{*}{2. Triggers and winds of change} & 2.1 Cultural awareness & 
\multirow{3}{.4\textwidth}{The theme summarizes the main factors that will form the value structure in future software organizations.} \\
& 2.2 Scaling agile & \\
& 2.3 Loyalty shift & \\
\multirow{3}{*}{3. Value misalignment} & 3.1 Objects of discrepancy & 
\multirow{3}{.4\textwidth}{The theme relates to the causes and effects of value misalignment between groups in software engineering organizations.} \\
& 3.2 Consequences & \\
& 3.3 Causes & \\
\hline
\end{tabular}
\caption{Identified themes and sub-themes.}
\label{table_themes}
\end{table}

\subsubsection{The dark side of agile} 
During the interviews, all of the participants mentioned the significant influence the agile transition had on the organizational values in software companies. Still, the comments they made were not solely positive. Several managers were of the opinion that the agile movement had grown too powerful, which, according to them, had unfavorable effects on organizational life.

\paragraph{Political correctness}
A number of participants stressed that the agile approaches introduce several drawbacks that the organizations were reluctant to discuss. These participants argued that the agile community, at its worst, promotes a cult-like behavior in which it was prohibited to question agile superiority. The top manager in Organization C said, somewhat despondently, that ``It has become difficult to challenge or even have an open discussion about the agile usefulness. If you don't embrace agile unconditionally, you're considered a dinosaur -- a dying organization.''

As an example, the majority of the interviewees acknowledged that delegating authority and responsibility increase the employees' engagement. Still, a few managers emphasized that delegating responsibility comes at a cost. In traditional organizations, if a project diverted from the initial plan, the software engineers could often be shielded from the reprimands that usually followed by their managers who had the overall responsibility. The managers thus had the role of identifying and compensating for ideas that did not work, and, in a way, acted as the adder of structure in areas that lacked clarity. Thereby, the managers could carry some of the burden of their employees' anxiety.

Agile organizations tend to lack the safety-net that is built-in into hierarchical organizations. If not managed correctly, that can add to the pressure and stress of the software engineers. The second line manager in Organization D stated that ``Teams or team members can become too committed. If something goes wrong, they feel personally responsible. They put immense pressure on themselves, which can lead to stress and burn-out.'' Thus, when creating autonomous and self-managing teams, the organization should add structure in the areas that were previously handled by the managers by default.

\paragraph{One ambiguous method to rule them all}
According to nearly all of the participants, agile was the only game in town when it came to software development methodology. Still, when the participants attempted to explain the core meaning of agile, their interpretations were quite varied. They attributed different meanings to the construct and, also, to what signified an agile organization. Several participants argued that the construct was too ambiguous, too open to interpretation, and too high-level, which meant that various parts of the organization had a different definition of what it means to be agile and was a source of confusion and misunderstandings. As a direct consequence, in Organization A, the differences in interpretation of the agile constructs had led to so much confusion and organizational discomfort that the management had chosen to ban the use of the word internally. The software engineer said that ``To facilitate communication, we talk about a team-based organization instead of an agile organization.''

\paragraph{Agile as trademark}
Some managers reported that when hiring developers, they regularly exaggerated their agile maturity level during interviews. Being agile had become a necessary trademark, a certificate, that indicated the lowest acceptable standard for software engineers. The relationship between employee and manager was, therefore, not off to a promising start, since the employee realized soon after arrival that the company's methods did not meet the promised expectations. As a consequence, the psychological contract between the hiring manager and the new employee was breached, which, naturally, had an adverse effect on organizational trust and set an unfavorable tone for the future relationship between employing managers and employees. 

\subsubsection{Triggers and winds of change}
As we previously have reported, agile approaches have significantly influenced the value structure of software engineering companies. However, during our interviews, several other factors surfaced that, according to the participants, are likely to considerably influence the values of software organizations in the future.

\paragraph{Cultural awareness}
A majority of the participants maintained that employees or managers rarely discussed the values that should govern behavior. Although the managers recognized the importance of shared values, they unwittingly prioritized and focused on more concrete issues, such as improving processes, methods, techniques, and tools. The human resources department initiated one of the rare occasions where the employees sat down to discuss values. The participants' opinions of such events, however, varied, and several participants indicated that the outcome (i.e. organizational core values) was too general and all-inclusive. These core values had, therefore, a minimal influence on organizational behavior. 
\begin{quote}
\emph{Identifying the core value could be a good thing. However, defining three words that should encompass the culture of several departments, where more than 10 000 employees work, will, without a doubt, result in them being so ubiquitous and vague that they, in the end, mean absolutely nothing.} (E2)
\end{quote}

Some participants explained that the purpose of the core values was not to alter organizational behavior, but rather to create a public profile that the company could exploit when communicating externally (i.e. a part of their organizational identity). There were, however, signals of increasing cultural awareness. Almost all managers reported that culture and values were climbing increasingly higher on the management teams' agendas. Management's attention to these concerns is also significantly higher nowadays than it was 15 years ago. The participants felt that this growing focus would undoubtedly benefit the software industry as a whole.

\paragraph{Scaling agile}
The introduction of the agile approaches emphasized teamwork, collaboration, and cooperation, and all of the participating companies in this study organized their work in teams. As the complexity and size of the software increased, many of the companies' products became too complicated for one team to manage. Therefore, inter-team collaboration had become increasingly important.

\begin{quote}
\emph{Before, we needed to get individual software engineers to collaborate. That was a challenge in itself. Now, we'll have to make groups of developers collaborate. That's much, much more challenging.} (E2)
\end{quote}

Three of the four organizations had started to implement a method for scaling agile. Of these, two had chosen to implement SAFe (Scaled Agile Framework), while one organization had purposely chosen to `reinvent the wheel.' The top manager in this organization argued that to facilitate multi-team collaboration, an organization must have common organizational values and a strong identity. The manager meant that the process of creating a tailored, scaled agile method itself strengthened shared values and formed the organization's identity. In his experience, without a strong identity and value structure, the system would be sub-optimized by having teams that are effective on their own, but cannot collaborate.

\paragraph{Loyalty shift}
A few managers stated that the loyalties of the software engineers had shifted over the past 10 years. Previously, the developers expressed loyalty towards the company or the company's products. This loyalty has decreased, and the reduction was, according to the managers, more apparent among software engineers compared to other types of engineers. They felt that a significant contributing factor was the increasingly influential role of teams in software-engineering organizations. The teams' cohesion had a notable effect on developers, who expressed greater loyalty to the teams' working processes than to the organization or its products.

Another development related to the job market that, according to several participants, has affected the behaviors and value structure of the large software companies is the growing shortage of software engineers. The more skilled developers have learned to exploit the market and maximize their profits by starting a consultant company of their own or by changing jobs frequently. A few managers pointed out that this development was somewhat incompatible with the trend of agile. The section manager in Organization D felt that ``It is becoming increasingly hard to keep the teams constant when the employee turnover rate increases.''

\subsubsection{Value misalignment}
The alignment of preferred organizational behavior differed significantly among the included organizations. In Organizations D and E, the narratives the interviewees used to portray their respective organizations overlapped; they had roughly the same view of what challenges the companies faced, what behaviors were prosperous, and the purpose and identity of their organizations. By contrast, the participants from Company A and C had a more fragmented view and their stories did not overlap to the same extent, instead providing an incoherent description of their respective organizations.

\paragraph{Objects of discrepancy}
All of the participants in Organizations D and E repeatedly expressed the opinion that autonomy and self-organization were desirable organizational traits, since they increased the software engineers' motivation. In Company A and C, the situation was more complicated, and the interviewed employees' beliefs were not as congruent. There was not, however, a clear-cut distinction between managers and developers. In both organizations, the interviewed developers' and section managers' values were aligned but differed from those of the upper management, (i.e. department and business unit managers). Although the upper managers acknowledged the trend towards increased autonomy and self-management, they seemed to hold conflicting feelings and from time to time argued against it during their interviews. Their statements of doubt gradually became more apparent during the discussion, as shown in a statement by C3: ``I think it is important that the right person makes the decisions. Sometimes it is evident that the engineers do not grasp the whole situation and can therefore not be expected to make a sound decision. They just don't know all of the relevant facts.''

The differences in opinion were also evident when discussing organizational role models. The section manager in Company C was inspired by ``the new generation of companies'' such as Google, Spotify, and Facebook, while the department manager stated that ``I am tired of employees comparing us to Google -- we are nothing like them.'' The department manager argued that it was unreasonable to compare their organization to Google; since Google produces software only, they can hire almost anyone they wish, and they do not have to battle an organizational culture that has been built up over 50 years. Drawing on this reasoning, the department manager felt that it was far from evident that methods that work well at Google should automatically function adequately in their organization as well.
 
Moreover, in the organizations with high misalignment (i.e. A and C), there were indications that the groups were not in agreement on what timeframe should be used when making decisions. As an example, the developer in Company C said that their organization frequently needed to ``live to fight another day,'' meaning that they were forced to make decisions within a short timeframe; in practice, this meant lowering the software quality by ignoring reusability and skipping automatic unit tests. The developers argued that optimizing for the near future is, at least to some degree, at odds with the ethical codes and the professional identity of the software engineering profession. During their education, the developers are taught to build for the future and to develop components that can withstand future trials. Continually adding to the technical department of a system thus made the software engineers uncomfortable, and reduced their engagement and motivation.

\paragraph{Consequences}
\label{lbl_consequences}
Our analysis indicated that between-group value misalignment affected the four performance factors of organizational effectiveness, conflict levels, between-group trust, and employee job satisfaction. The discrepancy of values within Organizations A and C led to tensions and conflicts between groups. Several participants suggested that in a misaligned organization, the employees did not fully understand what behaviors were expected of them when collaborating with employees from groups other than their own. This raised feelings of insecurity and adversely affected employees' job satisfaction.

In addition to not knowing how to act themselves, value misalignment led to the employees being unsure of how other groups would react and behave, which eroded the trust between organizational groups. This created an unpredictable working environment in which the employees were reluctant to make decisions themselves, instead frequently delegating decisions hierarchically upwards, most often to their managers. That reduced the organizational efficiency by delaying or in some cases halting the development process. The interviewed developer in Organization D explained that ``Often you cannot get a decision on a matter right away. So in order to not waste any time, one has to keep on working as if a decision has already been made. To do that with some confidence, one needs to be able to predict the decision in advance.''

Moreover, there were indications that the employees in Organization A and C were more inclined to focus on maximizing the performance of their team than the organization as a whole. As an example, there were notable differences in how the organizations described their goals. The participants from Organization D and E highlighted the organizational goals and how they, or their unit, could contribute to them. The employees of A and C, however, were more likely to emphasize low-level goals related to their group, whereas the company's overall goals were not typically mentioned. Their reasoning was the opposite when discussing organizational improvements. In this case, the employees in organization A and C often addressed issues outside of their control while the participants from D and E presented improvements they, themselves, could manage.

Another distinction was how the interviewees portrayed the relationship between themselves and the organization. The interviewees from D and E often identified themselves with the entire organization using the pronouns \emph{we} or \emph{our}, ``Our most important goal next year is to increase our sales.'' (D1). They saw themselves and their group as a part of a collective to which they belonged and contributed. The interviewees of A and C, on the other hand, were more inclined to distance themselves from the other groups, as illustrated by the following statement from A2: ``The products are already state-of-the-art, it is now up to the sales department. They need to step it up.''

\paragraph{Causes}
A distinction that separated the top managers in A and C from their colleagues in D and E was that the participants from the former argued that key aspects of the agile approaches, for example, autonomy and self-organization, had already been tried before. They considered agility to be a trend, something in passing that one did not need to focus on very much. The business unit manager in organization A stated that ``We [the company] tried self-organization 25 years ago, and it didn't work then. These things seem to go in circles.''

These managers thus saw no reasons to challenge their beliefs and, consequently, felt no sense of urgency to change them. The interviewed software engineers and the section managers, on the other hand, communicated that a transition to a more agile approach was a necessity. As a consequence, the adoption of agile approaches added to the differences in values between groups in organization A and C. By contrast, in organization D and E where the agile adoption process had a more company-wide acceptance, the changes reduced the differences in values.

A few participants argued that it was natural for large organizations to be misaligned during a transitional phase and that, sooner or later, the different organizational units would adopt the new values. The department manager at Organization D had a somewhat pessimistic mindset and stated that ``When it comes to such fundamental concerns, some people do not change their minds, but they eventually get replaced by someone else.''

As we previously reported, several of the participants argued that organizational values seldom surface as a topic in everyday operations, in which the purpose of a discussion is most often to solve an immediate problem or settle a heated discussion. If a conflict of interest arises, the involved employees usually aim for the short-term goal (i.e. to solve the current issue pragmatically). They rarely strive to identify a conflict's underlying cause, which may or may not be a misalignment of values. The reluctance to openly discuss values thus concealed the value misalignment, making the phenomenon harder to detect. One participant in Organization E said that ``Talking about values falls into the same category as talking about feelings. Engineers, in general, are rather uncomfortable with that and it is definitely not part of the engineering culture.''

\section{Part II - Testing the effects of value misalignment}
\label{section_study_b}
In the second part of the mixed model study, we statistically tested whether value misalignment affected the four performance factors (effectiveness, trust, conflicts, and job satisfaction) that we identified in the qualitative analysis (see Section \ref{lbl_consequences}).

\subsection{Method}
Since we aimed to verify the qualitative findings statistically, we chose a quantitative research approach with questionnaires.

\subsubsection{Sample}
We collected questionnaire data from the seven organizations described in Section~\ref{company_overview}. As is shown in Table~\ref{table_respondents}, 184 employees working in 27 different teams participated. In this study, we define a team as group of people working together to complete a task, job, or project. The majority of all respondents worked in Sweden, but a few (~10\%) were based in India. We did not collect background information from the respondents; however, since we conducted the data collection ourselves, we determined that 20\% were women. 

The participating teams were selected on the grounds of convenience~\cite{etikan2016comparison}. The upper management selected which teams in their organization we were allowed to contact. They all stated that the selection was based on availability and workload. We contacted the teams via e-mail and included the ones that first responded. Only a few teams declined to participate in the study. These stressed that they needed to focus on other, more pressing, matters.

\begin{table}
\centering
\begin{tabular}{ l l l l l}
\hline
Org. Id. & Man. teams & Dev. teams & HR teams & Total \\ \hline
A &     12 (2)  & 10 (2)    & 10 (1)    & 32 (5) \\
B &     18 (2)  & 17 (2)    & 0         & 35 (4) \\
C &     8 (2)   & 14 (2)    & 0         & 22 (4) \\ 
D &     15 (2)  & 16 (2)    & 0         & 31 (4) \\ 
E &     11 (2)  & 12 (2)    & 0         & 23 (4) \\
F &     13 (2)  & 6 (1)     & 0         & 19 (3) \\
G &     8 (1)   & 14 (2)    & 0         & 22 (3) \\
Sum: &  85 (13) & 89 (13)   & 10 (1)    & 184 (27) \\ \hline
\end{tabular}
\caption{Overview of respondents (teams) per organization.}
\label{table_respondents}
\end{table}

\subsubsection{Materials}
The result of the qualitative inquiry (see Section \ref{lbl_consequences}) indicated that alignment of values between organizational groups positively affected software companies' overall effectiveness, between group trust, level of conflicts and the employees' job satisfaction.

The study's intent was to gain general insights as to the significance of the \emph{between-group value misalignment} construct. We wished to compare the group-level construct with a similar individual-level construct that previously had been used to measure value congruence in organizations. One construct that met these criteria was \emph{value strength}, which we chose to include in our analysis. The \emph{value strength} construct provides a measurement of overall agreement among individual employees (i.e. not agreement between groups) regarding values in the organization~\cite{schneider2002climate}. Research concerning organizational climate has utilized and proven the significance of this type of concept~\cite{schneider2002climate}.


We measured the constructs using self-assessment questionnaires. However, in an attempt to triangulate and add support to the validity of the constructs of \emph{effectiveness} and \emph{job satisfaction}, we collected `real' project data (i.e. \emph{delivery success} and \emph{employee turnover}) from four of the seven participating organizations.

\begin{figure}
\centering
\includegraphics[width=0.75\textwidth]{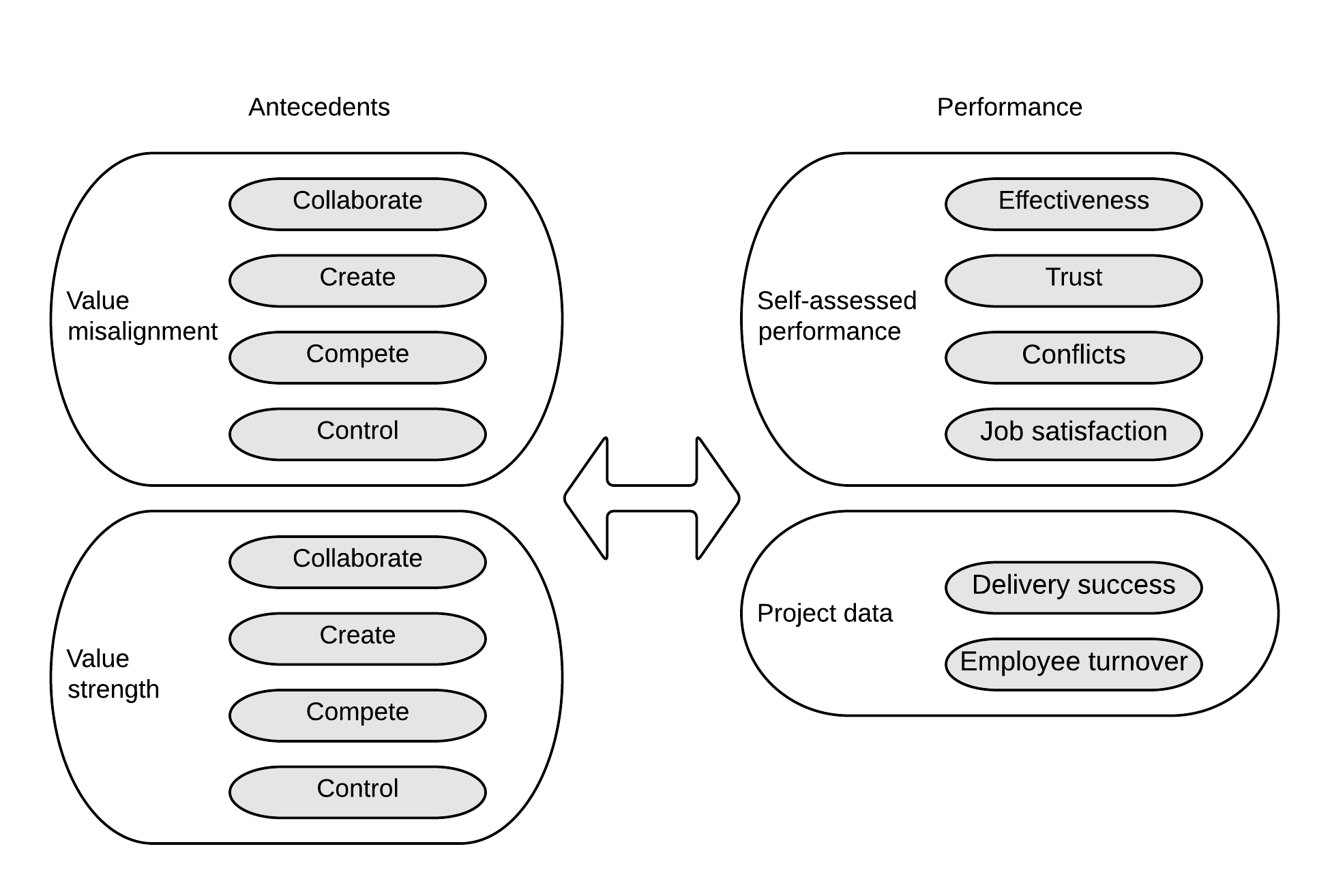}
\caption{The figure presents an overview of the constructs used in the study.}
\label{fig:overview_constructs}
\end{figure}

An overview of the included constructs is shown in Figure~\ref{fig:overview_constructs}.

\paragraph{Questionnaire}
We aimed to utilize previously verified constructs rather than developing new scales and items.

\subparagraph{Between-group value misalignment} We operationalized organizational values using the organizational culture assessment instrument (OCAI)~\cite{cameron2011diagnosing}, which is based on the CVF~\cite{hartnell2011organizational, cameron2011diagnosing, denison1991organizational, quinn1983spatial}. The survey measured both the \emph{existing} values and the respondents' \emph{preferred} values that would help their organizations to achieve their highest aspirations. In our analysis, we used the \emph{preferred} values, since we sought the respondents' desired preferences and beliefs, not their perceptions of the current state of their organization.

We chose to use the CVF because it relates to the core values in the organization, is well reported in the literature~\cite{hartnell2011organizational}, and has previously been used in a software engineering setting~\cite{iivari2011relationship, iivari2007relationship}. Furthermore, other viable alternatives (for example, the organizational culture inventory~\cite{cooke1988behavioral}) have the disadvantage of including over 100 questions; we reasoned that such a time-consuming survey would have resulted in a significantly reduced response rated. The CVF characterizes organizational values on two dimensions (see Figure\ref{fig_cvf_figure}). As we previously stated, the first dimension represents a set of competing values indicating to what degree an organization emphasizes centralization and control over decentralization and flexibility. The second dimension is the degree to which the organization is oriented toward its internal environment and processes over its external environment and relationships with outside entities.

Several methods are presented in the literature that could be used to estimate alignment or congruence between groups; the choice of which method to use is therefore far from evident. Some of the viable alternatives are commonly used in multi-level analyses to justify aggregation of individual data points to a group. In this type of research, data should demonstrate both definite differences between groups and agreement within groups. As a result, in an organization with aligned values between groups, the respondents' reported values are not affected by their groups.

Two constructs commonly used to estimate group-level properties are intraclass correlation coefficient (ICC) and eta-squared corrected ($\eta^2(C)$)~\cite{bartko1976various, bliese1998group}. In a comparison study, Shieh~\cite{shieh2012comparison} argued that even though researchers and reviewers are familiar with, and almost reflexively demand, ICC, empirical evidence demonstrates that further improvement may be obtained by adopting the eta-squared estimation since it performs better for small values of ICC. Considering that the accumulated knowledge shows that magnitudes of ICC collected from industry data tend to be small, we chose to report eta-squared as an estimation for \emph{between-group value misalignment}. Still, for parts of our data, we also calculated the ICC values, which yielded similar results to eta-squared. We are therefore reasonably convinced that our choice of method does not pose a major threat to the study's validity.

In an attempt to add strength to the \emph{between-group value misalignment} measurements, we also assessed the CVF data using analysis of variance (ANOVA)~\cite{agresti2011categorical}. We conducted ANOVA analyses for each of the seven organizations, with the four CFV values as dependent variables and team belonging as the independent variable. Since ANOVA analysis indicates the existence of differences between groups, we expected the organizations with high \emph{between-group value misalignment} scores to have significant ANOVA results.

\subparagraph{Value strength} We followed the advice of Schneider et al.~\cite{schneider2002climate} on climate strength and operationalized the \emph{value strength} construct as the (sign inverted) standard deviation of the four reported organizational values in CVF. We have noted, however, that the use of standard deviation as an estimate for strength has been criticized, as it is a measurement of disagreement (i.e. the opposite of agreement). Still, we agree with Schneider's reasoning and argue that such difference is negligible for our purposes and thus calculated \emph{value strength} estimates for each of the four dimensions of the CVF, where a higher value indicated more value agreement.

Homogeneity statistic $r_{wg}$ would be an alternative construct but has several drawbacks~\cite{bliese2000within}; for example, it may overestimate the degree of agreement and result in values greater than one, which are difficult to interpret~\cite{zohar2005multilevel}.

\subparagraph{Performance measurements} The organizational \emph{effectiveness} was estimated based on four items suggested by Cohen et al.~\cite{cohen1996predictive}, and the internal \emph{trust} between teams was measured using four questions define by Huff and Kelley~\cite{huff2003levels}. We used four items extracted from work by Jehn and Manix~\cite{jehn2001dynamic} (task conflicts) and Friedman et al.~\cite{friedman2000goes} (personal conflicts) to estimate the \emph{conflicts} between teams. Finally, \emph{job satisfaction} was estimated based on the four questions suggested by Thompson and Phua~\cite{thompson2012brief}. The items for measurements are listed in Table~\ref{table_items}.

\begin{table}
\footnotesize
\setlength\extrarowheight{2pt}
\centering
\begin{tabular}{ L{.15\textwidth} L{.45\textwidth}} \hline
Construct & Question  \\ \hline
Effectiveness 1 & This organization does high quality work. \\
Effectiveness 2	& This organization is effective. \\
Effectiveness 3	& This organization delivers according to schedule. \\ 
Effectiveness 4	& This organization rarely has cost overruns. \\

Trust 1 &	In this organization, there is a high level of trust between units. \\
Trust 2 &	In this organization, employees have a great deal of trust for managers.  \\
Trust 3 &	If someone in this organization makes a promise, others will almost always trust that the person will do his or her best to keep the promise. \\
Trust 4 & Managers in this organization trust their employees to make good decisions. \\

Conflict 1 & In this organization, one party frequently undermines another. \\
Conflict 2 & Much `plotting' (conspiring) takes place `behind the scenes' in this organization. \\
Conflict 3 & In this organization, people from different units often disagree about how work should be conducted. \\
Conflict 4 & There are often conflicts of ideas in this organization. \\
Job satisfaction 1 & I find real enjoyment in my job. \\ 
Job satisfaction 2 & I like my job better than the average person. \\ 
Job satisfaction 3 & Most days I am enthusiastic about my job. \\ 
Job satisfaction 4 & I feel fairly well satisfied with my job. \\ \hline

\end{tabular}
\caption{The table shows the items used to compile the four variables; effectiveness, trust, conflict, and job satisfaction.}
\label{table_items}
\end{table}

The answers to all of these items were measured using a five-point Likert scale with the following options: `strongly agree', `agree', `neither agree nor disagree', `disagree', and `strongly disagree'.

\paragraph{Real project data} Four of the organizations reported project data. For the four most recently completed projects or deliveries, the top management of these companies was asked to report if the project was completed according to the original schedule and if the original cost estimates were held. We then summarized the values for each organization (positive response = 1, negative response = 0), giving this measurement a value range between zero and eight, which was reported as a percentage. The organization also reported their average yearly employee turnover in the percentage of employees for the previous two years.

\subsubsection{Procedure}
All data collection was conducted over nine weeks starting at the beginning of March 2018. The questionnaires were distributed to the teams by the first author at their weekly or monthly meetings, earning us an overall response rate of over 90\%.

Before completing the survey, the respondents were informed about the general purpose of the research, that it was anonymous, and that it was voluntary to participate. The researcher also informed that he would not share the raw or analyzed data with other researchers or with their respective management.

The researcher and the respondents also discussed the concept of \emph{organization} that was used in the survey, as it was critical that the respondents from each organization used the same interpretation. We defined the concept as the part of the organization that was known to the respondent but limited it to the part of the company that the participating top management team was responsible for (i.e. the units hierarchically below them).

For the vast majority of the respondents, it took less than 20 minutes to answer the questionnaire. No respondents completed the survey in less than six minutes and the longest anyone sat was 45 minutes. To increase the number of correctly completed questionnaires, the researcher encouraged the respondents to review their answers before handing in the questionnaires. Of the 184 respondents, more than 95\% completed the entire survey correctly. Incorrect answers were ignored in the analysis.

We also asked the participating teams if they wished to receive information about their results in relation to the overall mean of the other participating teams. All teams were interested in this feedback.

\subsubsection{Analysis}
We reported the relationship between the two constructs \emph{between-group value misalignment} and \emph{value strength}, and the four performance measurements using Pearson correlation coefficients. Parts of the results are also presented visually using a scatter-plot.
Before the analysis, we tested the internal consistency for the included constructs using Chronbach alpha, which ranged from 0.73 (for conflict) to 0.86 (for job satisfaction). These alpha values were acceptable~\cite{tavakol2011making} and aligned with measurements in previous research~\cite{lenberg2018safety, jehn2001dynamic, huff2003levels, friedman2000goes}.
We also tested if the acquired information fulfilled the parametric statistical tests assumptions. First, we inspected the histogram, box-plot, and descriptive data for all constructs and confirmed that no outliers existed. Then we tested whether the variables were approximately normally distributed using the Shapiro-Wilk test. Finally, we tested the homogeneity of variances assumptions using the Levene test. All constructs met the required assumptions (i.e. none provided a significant result, which indicates that parametric tests are applicable).

We conducted this analysis using SPSS Version 24.

\subsection{Results}
The mean and standard deviation for the included constructs are presented in Table~\ref{res_mean_stddev}. The value range for the performance measurements are 1.0-5.0, where a higher value indicates a better result (i.e. more effective and fewer conflicts). The \emph{project success} construct reports the rate of success for the organization's previous four projects/deliveries, while the \emph{turnover} construct is the turnover as a percentage of the total number of employees. The table is sorted by \emph{effectiveness}, starting with the least effective organization. The table indicates a significant difference between the least effective organization and the most effective organization. The respondents in Organization G report, on average, 70\% higher effectiveness than employees in Organization C. Furthermore, the table indicates that collected project data (i.e. \emph{delivery success}) roughly follow the same pattern as self-assessed effectiveness, which, at least to some extent, adds support to the results of the self-assessed measurements. 

\begin{table}
\centering
\begin{tabular}{ccccccc}
\hline
\multicolumn{1}{|c|}{\multirow{2}{*}{Org. Id.}} & \multicolumn{4}{c|}{Performance} & \multicolumn{2}{c|}{Project data} \\ \cline{2-7} 
\multicolumn{1}{|c|}{} & \multicolumn{1}{c|}{Effectiveness} & \multicolumn{1}{c|}{Job satisfaction} & \multicolumn{1}{c|}{Conflict} & \multicolumn{1}{c|}{Trust} & \multicolumn{1}{c|}{Project Success} & \multicolumn{1}{c|}{Turnover} \\ \hline
C & 2.31(0.62) & 3.38(0.65) & 3.11(0.70) & 3.35(0.58) & 25\% & 10\% \\
A & 2.67(0.61) & 3.91(0.69) & 3.09(0.57) & 3.30(0.49) & 37\% & 12\% \\
B & 3.05(0.38) & 3.80(0.50) & 3.01(0.67) & 3.26(0.71) & 75\% & 24\% \\
F & 3.41(0.46) & 4.03(0.55) & 3.18(0.60) & 3.72(0.38) & - & - \\
E & 3.54(0.67) & 3.84(0.65) & 3.49(0.70) & 3.70(0.51) & 100\% & 3\% \\
D & 3.88(0.42) & 3.80(0.59) & 3.55(0.61) & 4.10(0.46) & 87\% & 10\% \\
G & 3.98(0.42) & 3.84(0.46) & 3.72(0.51) & 4.01(0.39) & - & - \\ \hline
\end{tabular}
\caption{The table presents the measured values (mean and standard deviation) of the four self-assessed performance measurements and the real project data.}
\label{res_mean_stddev}
\end{table}

The four calculated \emph{between-group value misalignment} scores (one for each dimension in the CVF) are presented in Table~\ref{res_misalignment}. As can be seen, C and A are the most misaligned organizations, indicated by relative high scores for the four \emph{between-group value misalignment} measurements. The employees in D and G, as a comparison, report significantly lower misalignment. The between-group misalignment for Organizations C and A is also strengthened by the ANOVA analyses, which reported statistically significant differences for two of the four dimensions. The ANOVA results for organization C were \emph{F(3,18) = 5.13, p = .010} (collaborate) and \emph{F(3,18) = 3.62, p = .033} (compete), while the results for Organization A were \emph{F(4,27) = 3.95, p = .012} (create) and \emph{F(4, 27) = 3.31, p = .025} (compete).


\begin{table}
\centering
\begin{tabular}{ccccccccc}
\hline
\multicolumn{1}{|c|}{\multirow{2}{*}{Org. Id.}} & \multicolumn{4}{c|}{Between-group value misalignment} & \multicolumn{4}{c|}{Value strength} \\ \cline{2-9} 
\multicolumn{1}{|c|}{} & \multicolumn{1}{c|}{Collaborate} & \multicolumn{1}{c|}{Create} & \multicolumn{1}{c|}{Compete} & \multicolumn{1}{c|}{Control} & \multicolumn{1}{c|}{Collaborate} & \multicolumn{1}{c|}{Create} & \multicolumn{1}{c|}{Compete} & \multicolumn{1}{c|}{Control} \\ \hline
C & 0.46 (A) & 0.12 & 0.37 (A) & 0.23 & -0.066 & -0.035 & -0.054 & -0.053 \\
A & 0.28 & 0.36 (A) & 0.32 (A) & 0.14 & -0.044 & -0.051 & -0.051 & -0.040 \\
B & 0.15 & 0.02 & 0.29 (A) & 0.14 & -0.068 & -0.054 & -0.058 & -0.072 \\
F & 0.03 & 0.05 & 0.02 & 0.09 & -0.065 & -0.057 & -0.064 & -0.051 \\
E & 0.17 & 0.13 & 0.09 & 0.13 & -0.053 & -0.043 & -0.056 & -0.041 \\
D & 0.05 & 0.04 & 0.05 & 0.15 & -0.065 & -0.052 & -0.069 & -0.036 \\
G & 0.04 & 0.06 & 0.03 & 0.04 & -0.059 & -0.060 & -0.069 & -0.058 \\ \hline
\end{tabular}
\caption{The table presents the \emph{between-group value misalignment} and \emph{value strength} scores for each of the four dimensions in the CVF framework (see Figure~\ref{fig_cvf_figure}). The (A) behind the \emph{between-group value misalignment} scores indicates that the ANOVA analysis reported a significant difference (p < .05) between groups for the corresponding CVF dimension.}
\label{res_misalignment}
\end{table}

\begin{figure}
\centering
\includegraphics[width=0.99\textwidth]{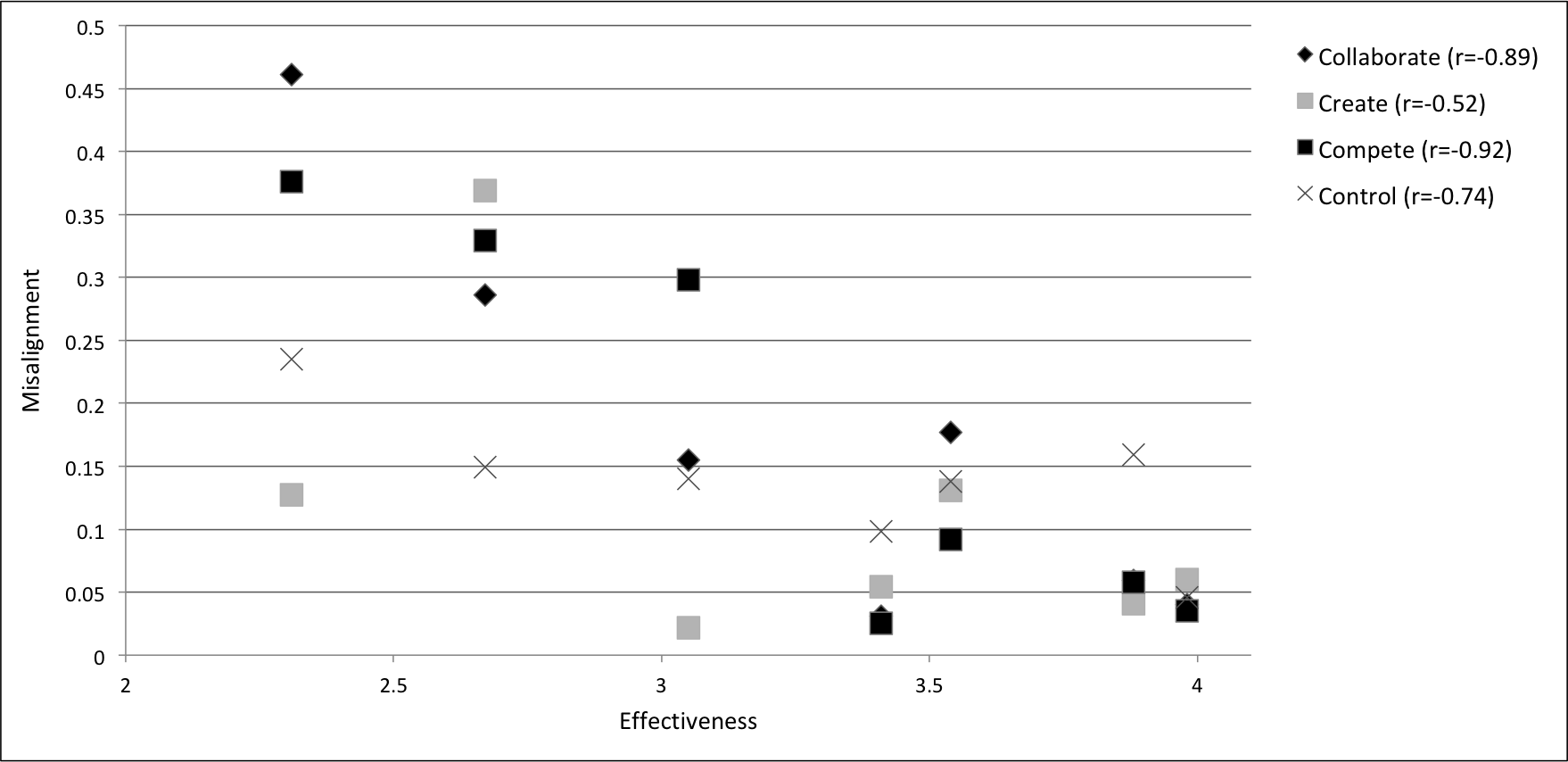}
\caption{The chart plots \emph{effectiveness} (X axis) against the four \emph{between-group value misalignment} measurements (Y axis).}
\label{fig:chart_corr}
\end{figure}

Table~\ref{res_correlation} shows the Pearson correlation coefficients for the relationship between two constructs \emph{between-group value misalignment} and \emph{value strength}, and the four self-assessed \emph{performance} measurements. The data clearly indicate that \emph{between-group value misalignment} relates to \emph{performance}. This suggests that organizations with low between-group misalignment perform better than organizations with high between-group misalignment. Using the rule of thumb advised by Cohen~\cite{cohen1992power}, we deem the link to be strong considering that the correlation coefficients are mostly above 0.5~\footnote{We note that the correlation between \emph{job satisfaction} and \emph{create} is considerably weaker compared to the other coefficients. Unfortunately, we cannot, using our current data, give a satisfactory explanation for this and, therefore, suggest that this is explored in future studies.}. The relationship is also displayed visually in a scatter plot (see Figure~\ref{fig:chart_corr}) for the self-assessed \emph{effectiveness} construct. 

Moreover, the data indicate a considerably weaker link between \emph{value strength} and \emph{performance} since these coefficients, in general, are small and not coherent. Our data thus suggest that \emph{performance} has a stronger relation to \emph{between-group value misalignment} than to \emph{value strength}.

\begin{table}
\centering
\begin{tabular}{c|cccccccc}
\cline{2-9}
 & \multicolumn{4}{c|}{Between-team value misalignment} & \multicolumn{4}{c|}{Value strength} \\ \cline{2-9} 
 & \multicolumn{1}{c|}{Collaborate} & \multicolumn{1}{c|}{Create} & \multicolumn{1}{c|}{Compete} & \multicolumn{1}{c|}{Control} & \multicolumn{1}{c|}{Collaborate} & \multicolumn{1}{c|}{Create} & \multicolumn{1}{c|}{Compete} & \multicolumn{1}{c|}{Control} \\ \hline
\multicolumn{1}{c}{Effectiveness}     & -0.89* & -0.52  & -0.92* & -0.74 & 0.10   & 0.64  & 0.85* & -0.15 \\ 
\multicolumn{1}{c}{Job satisfaction}  & -0.77* & 0.05 & -0.60  & -0.75 & -0.33  & 0.78* & 0.30 & -0.14 \\ 
\multicolumn{1}{c}{Conflict}          & -0.54  & -0.30  & -0.76* & -0.52 & -0.08  & 0.28  & 0.70 & -0.33 \\ 
\multicolumn{1}{c}{Trust}             & -0.71  & -0.46  & -0.89* & -0.51 & 0.14   & 0.41  & 0.88* & -0.38 \\ \hline 
\end{tabular}
\caption{Pearson correlation coefficients for the performance measurements and the two alignment constructs \emph{between-group value misalignment} and \emph{value strength}. The ´*' sign indicates a significant correlation at 0.05 level.}
\label{res_correlation}
\end{table}

\section{Discussion}
\label{section_discussion}
The primary objective of this study was to \emph{examine how discrepancies in values between organizational groups affect software companies' performance}. As a secondary objective, we sought to \emph{gain general insights into organizational values and how they affect behaviors and performance in software companies}. We used a mixed research design to meet these objectives. First, we collected qualitative data by interviewing 14 employees (software engineers (\textit{N} = 4) and managers (\textit{N} = 10)) working in four different software engineering organizations and processed the data using thematic analysis. We then conducted a quantitative survey of seven departments in six companies (\textit{N} = 184) to test the effects of value misalignment statically.

Regarding the primary objective, the results of our combined sub-studies show that, for our data, between-group value misalignment (a group-level construct) significantly affects software companies' performance. Organizations with low between-group value misalignment levels reported higher performance than those with more elevated levels. Aligned companies were more effective, more satisfied, had higher trust, and fewer conflicts. A similar, however individual-level construct that has been previously used to estimate value congruence in organizations is value strength~\cite{schneider2002climate}. Our analysis revealed that organizational performance had a stronger relation to between-group value misalignment than to value strength, indicating that between-group value misalignment is a more critical construct.

Moreover, our qualitative analysis suggests that diverse organizational values (i.e. high between-group value misalignment) create an unpredictable work environment, meaning that the employees cannot make an educated guess about future events. This creates a feeling of uncertainty, which raises stress levels among employees and lowers their sense of empowerment and autonomy. In contrast, companies in which the organizational values are aligned and known, create conditions for a stable and predictable working environment with fewer insecurities.

According to our findings, several factors contribute to the misalignment of values in software engineering organizations. First, a prerequisite for aligned values is an open dialogue, without which values are bound to diverge. However, our results suggest that shared values are seldom discussed in software organizations. Such dialogues tend to make software engineers feel uncomfortable and are not natural occurrences in organizational life. Since conversations about values seldom surface in everyday settings, employees learn acceptable behavior and shared values by studying others' behavior and reasoning. Drawing a parallel to social norms, one could argue that the values in software engineering organizations are descriptive rather than injunctive~\cite{cialdini1990focus}.

Secondly, since the agile construct is relatively elusive, each group's interpretations were biased by their respective prior understanding, experiences, and purposes. That has been recognized in our previous work~\cite{lenberg2018used}, as well as by other researchers~\cite{weyrauch2006we, laanti2013definitions, dikert2016challenges}. The results from this study suggest that a transition to an agile approach could, if successful, facilitate the normalization of organizational values and thus reduce between-group value misalignment, which, in turn, increases performance. However, if the meaning of the agile construct is not clarified and made common within the organization, an agile introduction could instead increase between-group value misalignment.

Thirdly, it naturally takes time to change organizational values. For a company to implement agile methods and processes is a swift change compared to a fundamental transformation of the organizational value structure. Companies can expect, and must take into account, that various organizational groups adopt new values at different paces, which at least temporarily increases between-group value misalignment. 

Regarding the secondary objective (i.e. to gain general insights as to organizational values and how they affect behaviors and performance in software companies), our study confirms the significant influence that the agile transition has had, and continues to have, on organizational values in software companies. By showing their usefulness in term of increased effectiveness, agile approaches have opened up alternative ways of organizing work by moving away from, and questioning, traditional Tayloristic principles and value foundations.

Our findings, however, indicate that the agile transition has not been a blessing only and that it should not be considered a `silver bullet'. As was reported by both the interviewed software engineers and the managers, the agile community has grown overly powerful and, at its worst, has created organizational values that prohibit questioning of the alleged agile superiority. To some extent, we acknowledge that this cult-like behavior was necessary when, at the turn of the century, the agile transition began to penetrate the norms and values foundations that then prevailed. Nevertheless, the agile community is no longer an underdog and its usefulness has strong empirical support~\cite{serrador2015does}. Such behaviors are no longer necessary and could, quite unnecessarily, contribute to a culture of silence that is likely to harm organizations.

As an example, it is commonly known that the agile organizations foster employee commitment and engagement. However, according to our interviews, agile advocators sometimes fail to recognize the fact that over-commitment and engagement may lead to stress and, potentially, burn out. In traditional organizations, the manager could, to a certain extent, carry the anxiety of his or her employees and thereby reduce the pressure. Agile organizations with autonomous teams, in which the managers have a less prominent role, should therefore replace the managers' responsibilities with additional structures or processes. Otherwise, the individual responsibility that comes with commitment might be too much to handle. In a culture of silence, where the agile way of working is considered flawless, the organization may neglect to implement such structures, resulting in excessive stress and reduced well-being.

We believe that our study contributes in several ways. In general, it adds support to previous research on alignment by reinforcing its importance in software engineering organizations and confirming the relationship between value alignment and performance~\cite{lynn2007literature}. Our work also extends existing alignment theory by providing initial empirical support for a between-group value misalignment theory. Such group-level theory is valuable for software engineering organizations in particular, since they almost exclusively organize their work in groups or teams. The software industry is currently transitioning to the use of scaled agile methods~\cite{dikert2016challenges} and additional insights as to which factors leverage inter-group collaboration are thus crucial.

The theory also provides additional or alternative explanations for the proven success of agile approaches. For instance, drawing on alignment theory, an agile transition contributes exclusively by forcing an organization to consider and reflect on their current value structure. Our results support this since in two of the participating organizations, the agile adoption processes had a harmonizing effects on organizations' internal values, thereby reducing misalignment and improving performance.

Moreover, our study contributes by providing additional general insights into organizational values in software engineering organizations. This is important because the current research on organizational values in software engineering has focused on exploring the fit between specific values and agile approaches~\cite{chandra2010identifying, hamid2015factors, dikert2016challenges}. Notably, we think it is essential to acknowledge the adverse effects that agile approaches can have on the value foundation. The research on agile has, so far, been somewhat optimistic and prone to report its strengths rather than its weaknesses~\cite{Barlow2011OverviewAG}. We believe that our findings can contribute to a more nuanced understanding of the agile concept.

Finally, our efforts have provided encouraging, yet initial results in an important research area. Given the relevance of teamwork and inter-team collaboration in software companies, we believe that it is important that between-group value misalignment is further, and more profoundly, explored in the software engineering context. Even if our findings indicate that there is a link between performance and between-group value misalignment, we have reason to presume that the relationship is not simplistic and linear. For example, we expect that organizations with virtually no misalignment create static psychosocial environments with few or no disagreements that stimulate creativity, improvement, and innovation. Drawing on the general agreement among scholars as to the relationship between tension levels and performance~\cite{de2006too, jehn1995multimethod, van1994optimising}, we suggest that between-group value misalignment has an inverted U-shaped relation to performance. At low and high levels of value misalignment, organizations are less effective than at moderate levels. Still, such a hypothesis, of course, requires empirical support from future studies.

\subsection{Limitations}
The study has several limitations. In the following sections, we discuss its internal validity, external validity, and content validity.

\paragraph{Internal validity}
Our survey included nearly 200 employees. Still, we used multi-level analysis, and base the conclusions on an analysis of aggregated constructs. We recognize that seven organizations are at the lower end of the scale, which we consider as a significant threat to internal validity. We thus regard our findings as preliminary.

We acknowledge that we had limited control of the selection process for both parts of the study, which could potentially affect the validity. Still, the upper management of each organization (which selected the teams we were permitted to contact) had little to gain from deluding us and seemed genuinely interested in providing an accurate view of their organizations. We therefore deem this particular threat to be limited.

Finally, a single researcher conducted all interviews and significant parts of the analysis. Throughout the study, all steps and phases have, however, were thoroughly discussed and reviewed by all authors as well as external experts. Still, in qualitative research, the researchers are instruments and are the most significant threat to qualitative research~\cite{poggenpoel2003researcher}. Even if we have taken measures to reduce the effects of researcher bias, we acknowledge that it has indeed colored the result.

\paragraph{External validity}
The survey and interviews were conducted in a limited number of companies located in Sweden, which threatens the generalizability of our findings. Still, the size of the unit limited the number of respondents, and the response rate of almost 90\% is acceptable.

\paragraph{Construct validity}
The representativeness of the survey measurements, which we estimated with self-assessment, is a threat to validity. In an attempt to raise the validity by triangulating the data (i.e. use additional data sources), we collected real project data. We recognize, however, a reduced resolution of the project data. Nonetheless, since we utilized items and scales that have been previously used and estimated the internal consistency, we believe that we can justify the use of the constructs in the study and thus rate this threat as moderate and acceptable.

We chose thematic analysis as the analysis method for the qualitative data. We acknowledge that there are other viable options, such as interpretive phenomenological analysis~\cite{smith2004reflecting}, content analysis~\cite{hsieh2005three}, or grounded theory~\cite{martin1986grounded}.

\section{Conclusions}
\label{section_conclusion}
This study reinforces the importance of considering organizational values in software companies. Empirical data collected from seven organizations indicated that value misalignment between groups is related to organizational performance. The aligned companies were more effective and more satisfied and had higher trust and fewer conflicts.

These findings can help to explain why some companies are more efficient than others and, thus, give initial direction to interventions addressing organizational challenges. Our results, for example, suggest that agile transitions have, in successful cases, forced the companies to clarify and evaluate their current organizational values. This has had a harmonizing effect on internal value structures, thereby reducing the misalignment and, in turn, improving performance.

Our efforts also provide more general insights on values in software engineering organizations. In particular, we emphasize the adverse effects that agile approaches can have on the organizational values foundation and hope that our findings can contribute to a more nuanced understanding of the agile concept.
\section*{Acknowledgments}
We acknowledge the support of Swedish Armed Forces, Swedish Defense Materiel Administration and Swedish Governmental Agency for Innovation Systems (VINNOVA) in the project number 2017-04874.
\bibliography{wileyNJD-AMA}

\section*{Author biography}


\begin{biography}
{\includegraphics[width=0.7in,keepaspectratio]{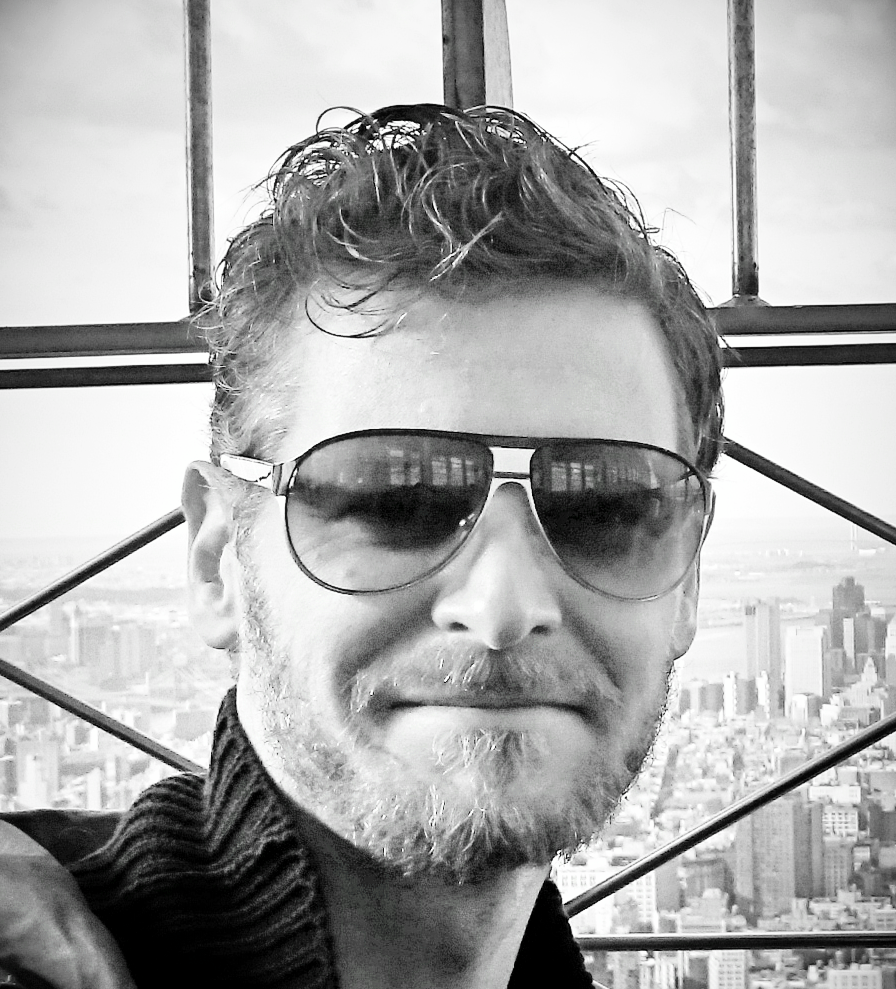}}
{\textbf{Per Lenberg} is an industrial PhD student in Software Engineering at Chalmers University of Technology in Gothenburg, Sweden. He has a M.Sc of Computer Science and Engineering from that same university and a Bachelor degree in Psychology from the University of Gothenburg. He has worked for more than 15 years in the software industry.
 
  .}
\end{biography}

\begin{biography}
{\includegraphics[width=0.7in,keepaspectratio]{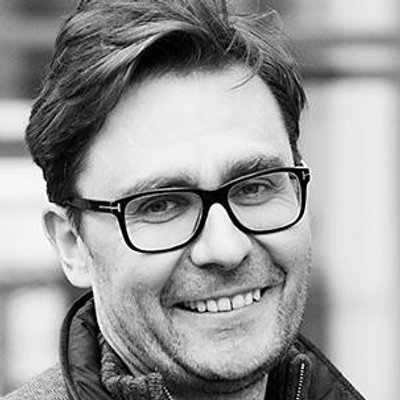}}
{\textbf{Robert Feldt} is a professor of Software Engineering at Chalmers University of Technology, Sweden and at Blekinge Institute of Technology, Sweden. He has also worked as an IT, AI and software consultant for more than 25 years. His research interests include human-centered software engineering, software testing and verification and validation, automated software engineering, requirements engineering and user experience. He has published more than 120 scientific papers. Most of his research is empirical and conducted in close collaboration with industry. He received a Ph.D. (Techn. Dr.) in computer engineering from Chalmers University of Technology in 2002.}
\end{biography}

\begin{biography}
{\includegraphics[width=0.7in,keepaspectratio]{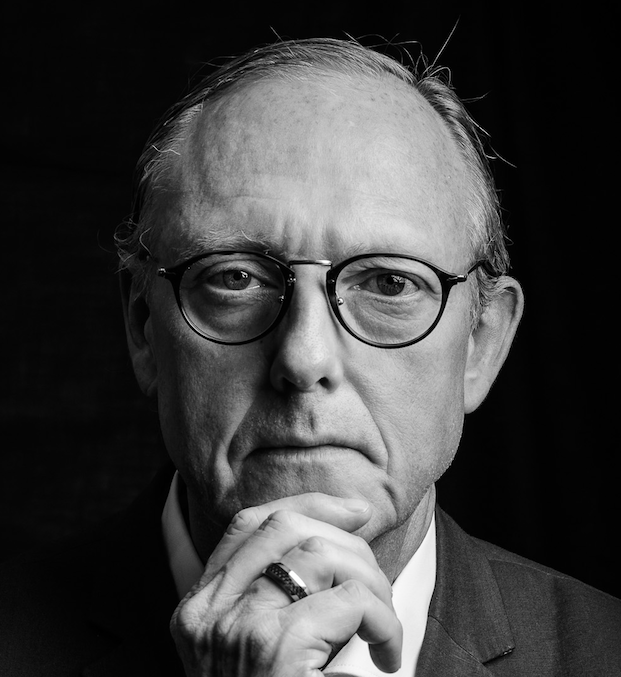}}
{\textbf{Lars G{\"o}ran Wallgren Tengberg} research interest encompasses the psychosocial work environment regarding leadership, stress, work characteristics with focus on work motivation, preferably within knowledge-intensive organizations. His doctoral thesis combines two key areas of occupational and organizational psychology, psychosocial work environment and work motivation, among IT consultants. He has a long and extensive experience of consultative work and development initiatives in the field of Information and Communications Technology (ICT).}
\end{biography}

\end{document}